\documentstyle[12pt]{article}

\begin{document}
\renewcommand{\thefootnote}{\fnsymbol{footnote}}
\begin{titlepage}
\begin{flushright}
HU-SEFT R 1996-05\\
IJS-TP-96/3\\
1996\\
\end{flushright}
\vspace{.5cm}
\begin{center}
{\Large \bf Long distance contribution to 
$K^{+} \rightarrow \pi^{+} \nu {\bar \nu} $ decay and 
$O(p^{4})$ terms in CHPT\\}
\vspace{.5cm}
{\large \bf S. Fajfer \\}
\vspace{.5cm}
{\it ``J. Stefan'' Institute, 
Jamova 39, p.p. 100, 
61111 Ljubljana, Slovenia\\}
\vspace{.5cm}
and \\
\vspace{.5cm}
{\it Research Institute for High Energy Physics, University of Helsinki, 
P.O.Box 9 (Siltavuorenpenger 20 C), FIN-00014 Helsinki, Finland \\}

\end{center}
\centerline{\large \bf ABSTRACT}
\vspace{0.5cm}
We calculate the long distance contribution to 
$K^{+} \rightarrow \pi^{+} \nu {\bar \nu} $  using chiral 
perturbation theory. The leading contribution comes from  $O(p^{4})$  
tree terms. The branching ratio of the $O(p^{4})$ long distance contribution 
is found to be of order $10^{-3}$ smaller than the short distance 
contribution.\\

\end{titlepage}
\renewcommand{\thefootnote}{\arabic{footnote}}
\setcounter{footnote}{0}
\vspace{.5cm}

{\bf 1 Introduction}\\

Chiral perturbation theory (CHPT) 
has been applied in the analysis of long distance effects in the 
$K^{+} \rightarrow \pi^{+} \nu {\bar \nu} $ decay  \cite{LW,GHL}.  
The observation of $K^{+} \rightarrow \pi^{+} \nu {\bar \nu} $ decay  
seems possible in the near future \cite{RW}. An upper limit 
on the branching ratio for the $K^{+} \rightarrow \pi^{+} \nu {\bar \nu} $ 
decay  
has been set at $2.4\times 10^{-9}$ $(90\% C.L.)$ \cite{BNL}.
The short distance loop diagrams dominate, due to the explicit 
dependence of the heavy top 
quark mass \cite{IL}.  
Sizable contributions come also from 
internal charm quark exchanges \cite{IL,EH}. 
The QCD corrections to this decay have been calculated 
in the leading logaritmic approximatation
\cite{EH,DDG,BBH}. 
Calculated next-to-leading QCD corrections reduced considerably 
the theoretical uncertainty 
due to the ambiguity in the choice of the renormalization scale present in 
the leading order expression \cite{BB,BB1,BLO}.
 
The 
$K^{+}(k) \rightarrow \pi^{+} (p) \nu (p_{\nu}){\bar \nu} 
(p_{\bar \nu})$ decay amplitude               
can be written in the form
\begin{eqnarray}
{\cal M} (K^{+} \rightarrow \pi^{+} \nu  {\bar \nu}) & = & 
\frac{G_F}{{\sqrt 2}} \frac{\alpha f_+}{2 \pi sin^2 \theta_W} 
[V_{ts}^{*} V_{td} \xi_t (m_t^2/M_W^2)  \nonumber\\
 +  V_{cs}^{*} V_{cd} \xi_c (m_c^2/M_W^2) & + & 
V_{us}^{*} V_{ud} \xi_{LD}]
(k + p)^{\mu}
{\bar u}(p_{\nu}) \gamma_{\mu} (1- \gamma_5) v (p_{{\bar \nu}}), 
\label{i1}
\end{eqnarray}
where $G_F$ denotes the Fermi constant, $\theta_W$ is the weak mixing angle, 
$\alpha$ is the fine structure constant, 
$f_+$ is the form factor in ${\bar K}^0 \to\pi^+ e {\bar \nu}_e$ decay 
and $V_{ij}$ stands for  $i \to j$ element of the CKM matrix. The functions 
$\xi (x_q)$ arise from the short 
distance contribution to 
$K^{+} \rightarrow \pi^{+} \nu {\bar \nu} $ decay \cite{EH,BB} and 
$\xi_{LD}$ from the long distance contribution \cite{LW,GHL,RS,HL}. 

In this paper we concentrate on  the long distance contribution to the 
part of amplitude $\xi_{LD}$,  
that arises from the time ordered product of the weak $\Delta S = 1$ 
effective Hamiltonian with the 
$Z^0$ neutral current. Due to the $\Delta I = 1/2$ rule \cite{LW,RS}, 
this contribution dominates the one 
coming from Feynman diagrams 
with two $W$ bosons. 

In CHPT the $O(p^2)$ tree level $K^+\to \pi^+ Z^0 \to \pi^+ \nu {\bar \nu}$ 
decay amplitude vanishes  \cite{LW,GHL}. 
The loop contributions were calculated in \cite{GHL}, 
leading to the branching ratio 
$BR(K^{+} \rightarrow \pi^{+} \nu {\bar \nu}) = 7.7 \times 10^{-18}$, 
roughly of order $10^{-7}$ smaller 
than that of the short distance contribution \cite{EH}.

We determine the long distance contribution, using the $O(p^4)$ 
strong and weak Lagrangians of CHPT. 
The number of counterterms 
in the effective weak Lagrangian of $O(p^4)$ is very large \cite{EKW,KMW,KMW1}.
However, it is possible to reduce them, using 
predictions for the weak Lagrangian 
in specific models [14-23]. 
%\cite{EPR2,EPR3,EPR4,EKW,IP,KMW,KMW1,SF0,BP}
We construct the weak Lagrangian 
of $O(p^4)$ relying  on the factorization approach.  

The paper is organized as follows: 
In Sect. 2 we discuss effective strong and weak  Lagrangians of 
$O(p^{4})$. 
Sect 3. contains the long distance contribution to the 
$K^+ \rightarrow \pi^+ \nu {\bar \nu} $ decay amplitude 
and the branching ratio.  
A short conclusion is presented in Sect. 4. \\

{\bf 2 Effective Lagrangians at $O(p^{4})$ in CHPT}\\

At the lowest order in momentum $O(p^{2})$ the strong chiral Lagrangian is 
given by 
\begin{eqnarray}
{\cal L}^{2}_{s} & = & \frac{f^{2}}{4}\{Tr ( D_{\mu}U^{\dag} D^{\mu}U)
+ Tr (\chi U^{\dag} + U \chi^{\dag})\} ,\label{e2}
\end{eqnarray}
where $ U = -\frac{i \sqrt{2}}{f} \phi$, 
$ f \simeq f_{\pi} = 0.093$ GeV is a pion decay constant and 
$\phi $ is a pseudoscalar meson matrix (see eg. \cite{EPR2,EPR3,GL}). 
The covariant derivative is given by 
$ D_{\mu}U = \partial_{\mu} U + i Ul_{\mu} -ir_{\mu} U$, $l_{\mu}$ 
and $r_{\mu}$ are external gauge field sources. 
The explicit chiral symmetry breaking induced by the 
electroweak currents of the standard model corresponds to the following 
choice:
\begin{eqnarray}
r_{\mu} & = & e Q[A_{\mu} - tan{\theta}_W Z_{\mu}]\label{e2h}
\end{eqnarray}
\begin{eqnarray}
l_{\mu} & = & e Q[A_{\mu} - tan {\theta}_W Z_{\mu}] + 
\frac{e}{sin \theta_W} Q_L^{(3)}Z_{\mu}  \nonumber\\
& + & \frac{e}{{\sqrt 2} sin \theta_W}[ Q_L^{(+)} W_{\mu}^{(+)} 
+ Q_{L}^{(-)}W_{\mu}^{(-)}]. 
\label{e2hh}
\end{eqnarray}
Here the $Q'$s are the electroweak matrices
\begin{eqnarray}
Q & = &\frac{1}{3} diag(2,-1,-1), 
\enspace Q_L^{(3)} = \frac{1}{2} diag(1,-1,-1) \label{e2hd}, 
\end{eqnarray}
\begin{eqnarray}
Q_L^{(+)} &  = & (Q_L^{(-)})^{\dag} = \pmatrix{
0 & V_{ud} & V_{us}\cr 
0 & 0 & 0\cr
0 & 0 & 0\cr}.  
\label{e2h2d}
\end{eqnarray}

The matrix $\chi = 2B[s(x) + i p(x)]$, with 
$s+ ip = {\cal M} $ 
$ = diag(m_u,m_d,m_s)$, 
takes into account the explicit breaking due to the quark masses 
in the underlying QCD Lagrangian \cite{GL,EGPR}. 
The matrix $\chi$ can be fixed by 
physical pseudoscalar masses to lowest order in the chiral expansion. 
The Lagrangian of $O(p^4)$ order in CHPT \cite{GL,EGPR} is given by 
\begin{eqnarray}
{\cal L}^{4}_{s} & = &L_1(Tr ( D_{\mu}U^{\dag} D^{\mu}U))^2 
+ L_2Tr (D_{\mu}U^{\dag} D_{\nu}U) Tr ( D^{\mu}U^{\dag} D^{\nu}U) 
\nonumber\\
& + & L_3 Tr (D_{\mu}U^{\dag} D^{\mu}U D_{\nu}U^{\dag} D^{\nu}U) 
+ L_4 Tr ( D_{\mu}U^{\dag} D^{\mu}U) Tr (\chi U^{\dag} + U \chi^{\dag})
\nonumber\\
& + &L_5 Tr ( D_{\mu}U^{\dag} D^{\mu}U(\chi U^{\dag} + U \chi^{\dag}))
+ L_6 (Tr (\chi U^{\dag} + U \chi^{\dag}))^2 \nonumber\\
& + &L_7 (Tr (\chi^{\dag}U -  \chi U^{\dag}))^2 
 +  L_8 Tr (\chi^{\dag} U \chi^{\dag}U + \chi U^{\dag} \chi U^{\dag})) 
\nonumber\\
& - i & L_9 Tr (F_R^{\mu \nu}  D_{\mu}U  D_{\nu}U^{\dag} + 
F_L^{\mu \nu}  D_{\mu}U^{\dag}  D_{\nu}U )
 + L_{10} Tr (U^{\dag}F_R^{\mu \nu} U F_{L \mu \nu}) \nonumber\\ 
& + & H_1 Tr (F_{R \mu \nu} F_R^{\mu \nu}+ F_{L \mu \nu} F_L^{\mu \nu}) 
+ H_2 Tr(\chi^{\dag} \chi)
\label{e4}
\end{eqnarray}
where $F_L^{\mu \nu} = \partial^{\mu} l^{\nu}  - \partial^{\nu} l^{\mu} 
-i [l^{\mu}, l^{\nu}]$  and $F_R^{\mu \nu} = \partial^{\mu} r^{\nu}  
- \partial^{\nu} r^{\mu} -i [r^{\mu}, r^{\nu}]$ and $L_1, \ldots L_{10}$ are 
ten real low-energy coupling constants, which completely determine the 
low-energy behaviour of 
pseudoscalar meson interactions to $O(p^4)$ ($H_1$, $H_2$ are of no physical 
significance). This $O(p^4)$ Lagrangian contains all possible terms which 
are allowed by chiral invariance. All the one loop divergences, which by 
power counting can only give rise to local $O(p^4)$ terms, are absorbed 
by suitable renormalizations of the $L_i$ and $H_{1,2}$ constants, as it was 
performed in \cite{GL}. Generally $L_i$ are divergent (except $L_3$, and 
$L_7$) and they depend on a renormalization scale $\mu$, which is not seen 
in observable quantities. 
The $H_i^r$ can be determined, although they are not accessible 
experimentally. In fact, once the $O(p^4)$ 
strong chiral Lagrangian  is chosen in the form of  \cite{GL,EGPR}, 
these couplings can be fixed by 
resonance exchange.

The authors of \cite{EGPR} 
calculated contributions of all low-lying resonances to the 
renormalized couplings $L_i^r$. Using the equations of 
motion for resonances, they found that the $L_i^r$, at 
the scale $\mu = M_{\rho}$, 
get contributions from vector, axial vector and scalar resonances.  
In the Table 1 we give the numerical values of  the $L_i^r$ and $H_1^r$, 
which we use later in our calculations. 
The Model $I$ in the Table 1 denotes the $L_i^r$ and $H_1^r$ from \cite{EGPR}. 

In our numerical calculations we use the values of the coupling 
constants $L_i^r$ and $H_i^r$ derived within the extended Nambu and 
Jona-Lasinio model \cite{BBR}. We concentrate on the 
fit $1$ and the fit $ 4$ of \cite{BBR}. The reasons are following: 
the fit 1 is obtained for the most favorable set of parameters required by the 
extended Nambu and Jona Lasinio model \cite{BBR}, while the fit 4 
was quite successful in explaining the experimental 
result for the $K^+ \to \pi \gamma^* \to \pi e^+ e^-$ 
weak counterterm \cite{SFZP}. 
The Model II (the fit 1 and the fit 4) in the Table 1 refers to the 
$L_i^r$ and $H_1^r$  from \cite{BBR}. 

The weak Lagrangian of $O(p^{2})$ in 
the chiral expansion has a unique form [14-18]
%\cite{EPR2,EPR3,EPR4,EKW,IP}
\begin{eqnarray}
{\cal L}_{w}^{2} & = & G_{8} f^{4} Tr (\lambda_{6} D_{\mu}U^{\dag} D^{\mu}U )
%J_{\mu} J^{\mu})
.\label{e1}
\end{eqnarray}
The chiral coupling  $G_{8}$ can be expressed using the well known constants 
in the Standard model
$G_{8}= \sqrt{\frac{1}{2}} G_{F} V_{ud} V_{us}^* g_{8}$.  
The coupling $|g_{8}| = 5.1$ was found analysing 
$K\rightarrow \pi \pi $ decay \cite{EKW,KMW1,IP}. 

The charged weak current to lowest order in chiral perturbation theory 
can be derived as \cite{EKW,IP,SF0}
\begin{eqnarray}
J_{\mu}^{(1)}|_{ij} & = & \frac{\delta S_{2}}{\delta (l_{\mu})_{ij}}. 
\label{c4}
\end{eqnarray}
$S_{2}$ is the action determined by ${\cal L}_{s}^{2}$ in (\ref{e2}), 
indeces $i,j$ denote $u,d,s$ flavor. 
The factorization model results in writing  
the weak Lagrangian (\ref{e1}) as \cite{EKW}
\begin{eqnarray}
{\cal L}_{w}^{2} & = & 4 G_{8} Tr (\lambda_{6} J_{\mu}^{(1)} J^{(1) \mu})
.\label{e1h}
\end{eqnarray}
The generalization on higher order terms in momentum leads to 
\cite{EKW,KMW,KMW1,IP,SF0} 
\begin{eqnarray}
J_{\mu}|_{ij} & = & J_{\mu}^{(1)}|_{ij} + 
J_{\mu}^{(3)}|_{ij} + \cdots \nonumber\\
& = & \frac{\delta S_{2}}{\delta (l_{\mu})_{ij}} 
+ \frac{\delta S_{4}}{\delta (l_{\mu})_{ij}} 
+ \cdots, 
\label{e4a}
\end{eqnarray}
where $S_4$ stands for the effective action at given order 
in momentum. The effective weak Lagrangian is then 
\begin{eqnarray}
{\cal L}_{w} & = & 4 G_{8} Tr (\lambda_{6} J_{\mu} J^{\mu})
.\label{e1hh}
\end{eqnarray}
There are other approches used to construct the 
weak Lagrangian at $O(p^4)$ \cite{EKW,EPR2,EPR3,EPR4,BP}.  
A certain confidence in the applicability of the factorization approach, 
we gain from  the analysis of the 
$K^+ \to \pi^+ \gamma^* \to \pi^+ e^+ e^-$ decay, which branching ratio 
is measured. 
In this decay, the finite part of the long 
distance contribution seems to be better described within the
factorization approach than within the weak deformation model \cite{SFZP}.
\\

{\bf 3 $K^+ \rightarrow \pi^{+} \nu \nu$ decay and  $O(p^{4})$  
contributions}\\

We write the 
$K^{+}(k) \rightarrow \pi^{+} (p) \nu (p_{\nu}){\bar \nu} 
(p_{\bar \nu})$ 
decay amplitude  in the following form 
\begin{eqnarray}
M (K^+ \to \pi^{+} \nu {\bar \nu} )
& = & \frac{e}{4 sin \theta_W cos \theta_W} M_{\mu} 
{\bar u}(p_{\nu}) \gamma_{\mu} (1  -  \gamma_5) v (p_{{\bar \nu}})
\label{ea1}
\end{eqnarray}
with 
\begin{eqnarray}
M_{\mu} & = &  g_+ (q^2) ( k + p )_{\mu} + g_- (q^2) ( k - p )_{\mu} 
\label{eb1}
\end{eqnarray}
The component along the momentum transfer $q_{\mu} = (k-p)_{\mu}$ 
gives no contribution to the unpolarized decay width. 
At the leading order $O(p^2)$ the sum of the pole and direct weak 
transition vanishes \cite{LW,GHL}. 

Using the strong Lagrangian given in (\ref{e4}),  
and the weak 
Lagrangian of $O(p^{2})$ from (\ref{e1}), we
obtain "indirect" (pole contribution) for the 
$K^+ (k) \to \pi^+ (p)  Z^0(\epsilon , q)$  decay, Fig. 1 (a) :
\begin{eqnarray} 
< \pi^+ Z^0 | {\cal L}_{pole}|K^+> 
& = &  
< \pi^+ Z^0 | {\cal L}^{4}_{s}|\pi^+>  
\frac{-1}{m_{K}^{2}  -  m_{\pi}^{2}}
< \pi^+ | {\cal L}_{w}^{2} | K^+>\nonumber\\
& + &  < \pi^+ | {\cal L}_{w}^{2} | K^+>
\frac{-1}{m_{\pi}^{2}  - m_{K}^{2}}
< K^+ Z^0 | {\cal L}^{4}_{s}|K^+>.
\label{pole1}
\end{eqnarray}
The direct weak transition is present also, Fig. 1 (b) 
\begin{equation}
< \pi^+ Z^0 | {\cal L}^{4}_{w}|K^+> 
= 4 G_{8} < \pi^+ Z^0 | 
Tr(\lambda_{6} \{ J_{\mu}^{(1)}, J^{(3)\mu}\})|K^+> 
\label{dir1}
\end{equation}  
with the weak current calculated using (\ref{e4a}). 
Neglecting terms of order $m_{\pi}^2/m_K^2$ in the decay 
amplitude $K^+ (k) \to \pi^+(p)  Z^0(\epsilon , q)$, 
we obtain for the indirect (pole) transition of $O(p^4)$ 

\begin{eqnarray}
< \pi^+ Z^0 | {\cal L}_{pole}|K^+> & = & 
4 i G_8 \epsilon \cdot (k + p) 
( L_4^r 8 m_K^2  + L_9^r q^2 )(s - t) 
\label{pole}
\end{eqnarray}
and for the direct weak transition of $O(p^4)$
\begin{eqnarray}
< \pi^+ Z^0 | {\cal L}_{dir}|K^+> & = & 
4 i G_8 \epsilon \cdot(k + p) 
[ - L_4^r 16 (s -t) m_K^2 + L_5^r 4(t - 2s)  m_K^2 \nonumber\\
& + & L_9^r 2(t - s) q^2 + L_{10}^r \frac{2}{3} t q^2 + 
H_1^r( \frac{4}{3} t -4 s) q^2 ]. 
\label{direct}
\end{eqnarray}
We use the notation
\begin{eqnarray}
s & \equiv & \frac{i e }{2 sin \theta_W cos \theta_W}, 
\enspace t \equiv ie tan \theta_W. 
\label{W1}
\end{eqnarray}

The authors of \cite{GHL} have calculated the finite part of the one 
loop amplitude. They found that there is a divergent part of $O(p^4)$. 
There is a standard procedure used to cancel out 
divergent contributions coming from the one loop effects of $O(p^4)$ 
\cite{EPR2,EPR3,GL}: one constructs the $O(p^4)$ counterterms in such a way,  
that divergences are reabsorbed in a 
renormalization of the various coupling constants. 
Therefore, we assume that the divergences 
coming from the loops of $O(p^4)$ \cite{GHL}
are cancelled out by divergent parts of the couplings of $O(p^4)$. 
It means that the contributions calculated 
in (\ref{pole}) and (\ref{direct}) 
determine the finite part of the cunterterms of $O(p^4)$. 
The $g_{+}^r(q^2)$ form factor is found from 
(\ref{pole}) and (\ref{direct}) to be
\begin{eqnarray} 
g_{+}^r(q^2)& = & \frac{2 e G_8}{m_Z^2sin \theta_W cos \theta_W} 
\{8 m_K^2 L_4^r (2 sin^2 \theta_W - 1)  + 8 m_K^2 L_5^r (sin^2 \theta_W - 1)
 \nonumber\\
& + & q^2 [L_9^r (2 sin^2 \theta_W - 1) + \frac{4}{3} L_{10}^r sin^2 \theta_W 
+ H_1^r ( \frac{8}{3}  sin^2 \theta_W -4)]\}.
\label{g+}
\end{eqnarray}
We neglect the logarithimic piece,  
since it was found to be very small \cite{GHL}. 

The size of the long distance contribution $\xi_{LD}$ defined in (\ref{i1}),  
can be estimated by taking $q^2 = 0$ in the form factor $g_+^r(q^2)$ 
in (\ref{g+}), and $f_+(0)= 1$ in (\ref{i1}). 
We obtain  the amplitude $\xi_{LD}(0) = -2.6 \times 10^{-5}$, using the 
couplings $L_i^r$ and $H_1^r$ from the Table 1 (Model I).
With $L_i^r$ and $H_1^r$ found in the fit $1$ (Model II, Table 1) , 
the amplitude becomes $\xi_{LD}(0) = -3.0 \times 10^{-5}$. 
With the couplings from the fit 4 (Model II, Table 1),  
the amplitude obtains the largest value $\xi_{LD}(0) = -3.2 \times 10^{-5}$.
These results are in agreement with 
the estimation made by Lu and Wise \cite{LW}. 
They  have noticed that the long distance contribution to 
$K^+ \to \pi^+ Z^0  \to \pi^+ \nu {\bar \nu}$ in CHPT might be 
of order $10^{-5}$. This should be compared with 
the charm quark short distance contribution $\xi_c \simeq 10^{-3}$.

The form factor $g_{+}^r(q^2)$ in (\ref{g+}) helps to calculate the 
decay width for $K^+ \to \pi^+ \nu {\bar \nu}$
\begin{eqnarray}
\Gamma(K^+ \to \pi^+ \nu {\bar \nu}) & = & 
C \int_{m_{\pi}}^{\frac{m_K^2 + m_{\pi}^2}{2 m_K}}  
|g_{+}^r(q^2)|^2 (E_{\pi}^2 - m_{\pi}^2)^{\frac{3}{2}} d E_{\pi},
\label{width}
\end{eqnarray}
with
\begin{eqnarray}
C & = & \frac{\alpha m_K}{24 \pi sin^2 \theta_W cos^2 \theta_W} .
\label{widthc}
\end{eqnarray}
Using the numerical values of $L_i^r$ $H_1^r$ from 
Table 1 (Model I), 
we calculate the branching ratio 
$BR(K^+ \to \pi^+ \nu {\bar \nu})_{LD}  = 0.17 \times 10^{-13}$.   

The couplings $L_i^r$, obtained in the fit $1$ (Model II, Table 1),  
lead to 
$BR(K^+ \to \pi^+ \nu {\bar \nu})_{LD}  = 0.29 \times 10^{-13}$, 
and the couplings $L_i^r$, from the fit $4$  (Model II, Table 1)  
give the largest, among calculated, branching ratio 
\begin{equation}
BR(K^+ \to \pi^+ \nu {\bar \nu})_{LD}  = 0.40 \times 10^{-13}.
\label{br4}
\end{equation} 
For comparison, we mention that the branching ratio 
coming from the short distance contribution is 
very close to $10^{-10}$ \cite{BB,BB1,GHL2}. \\

{\bf 4 Concluding Remarks}\\

The long distance contribution to 
the $K^+ \to \pi^+ Z^0  \to \pi^+ \nu {\bar \nu}$ decay is calculated 
using the $O(p^4)$ tree level of CHPT. 

The amplitude and the branching ratio depend on the 
model used for the $O(p^4)$ couplings 
$L_{i}$ and $H_{i}$ (the resonance exchange model 
\cite{EGPR} or the extended Nambu and Jona-Lasinio model 
\cite{BBR}). 

The branching ratio  can be as large as 
$BR(K^+ \to \pi^+ Z^0  \to \pi^+ \nu {\bar \nu})_{LD} = 
0.4 \times 10^{-13}$. This is still too small in 
comparison with the dominant top and charm quark
short distance contribution. Therefore, we support the 
suggestion of \cite{LW,RW,BB,BB1,BLO} that 
with the known mass of top quark a measurement of the 
branching ratio $BR(K^+ \to \pi^+ \nu {\bar \nu})$ 
would lead to a precise determination of the 
mixing angle $V_{td}$.\\

{\bf Acknowledgments} \\

The author thanks M. Chaichian for very useful discussions  
and his hospitality in Helsinki. 

This work was supported in part by the Ministery of 
Science and Technology of the 
republic of Slovenia. \\

\newpage
 
{\bf Figure Caption}\\

Fig. 1. Feynman diagrams which contribute to 
the $K^+ \to \pi^+ Z^0 $  vertex in CHPT. 
The square detones the weak interaction vertex of $O(p^2)$. 
The crossed circle (square) stand for the $O(p^4)$ strong (weak) 
vertex. 
\newpage
\begin{table}[h]
\begin{center}
\begin{tabular}{|c||r|r|r|}
\hline
$O(p^4)\enspace coup.$ & $Model\enspace I$ & 
$Model \enspace II (fit \enspace 1)$ &  $Model \enspace 
II (fit  \enspace 4)$\\
\hline\hline 
$L_5^r$ & $1.4 \times 10^{-3}$ & $1.6 \times 10^{-3}$& $1.7 \times 10^{-3}$ \\
$L_9^r$ & $6.9 \times 10^{-3}$ & $7.0 \times 10^{-3}$& $7.1 \times 10^{-3}$ \\
$L_{10}^r$ & $-6.0 \times 10^{-3}$ & $-5.9 \times 10^{-3}$& 
$-5.1 \times 10^{-3}$ \\
$H_1^r$ & $-7.0 \times 10^{-3}$ & $-4.7 \times 10^{-3}$& 
$-2.4 \times 10^{-3}$ \\
\hline
\end{tabular}
\end{center}
\caption{ The $L_i^r$ $i = 5,9,10$ and $H_{1}^r$ denote the $O(p^4)$ 
couplings calculated in the 
resonance exchange model of [25] (Model I) 
and the extended Nambu and Jona Lasinio model of [26] (Model II). 
The couplings coming from the fit 1 (fit 4) in [26] are presented 
in the second (third) column. (The $L_4^r= 0$ in both models.)  } 
\end{table}

\end{document}